\documentclass[
reprint,
nofootinbib,
nobibnotes,
amsmath,amssymb,
aps,
prd,
]{revtex4-2}

\usepackage{graphicx}
\usepackage{dcolumn}
\usepackage{bm}
\usepackage{url	}
\usepackage{float}

\usepackage{hyperref}
\usepackage{cleveref}
\usepackage[dvipsnames]{xcolor}

\hypersetup{
    colorlinks,
    citecolor=[rgb]{0.16,0.53,0.80},
    linkcolor=[rgb]{.88,.10,.25},
}

\newcommand*\aap{Astron.~Astrophys.}

\newcommand*\apjl{Astrophys.~J.~Lett}

\newcommand*\apjs{Astrophys.~J.~S}

\newcommand*\apss{Ap\&SS}

\newcommand*\jcap{J. Cosmol. Astropart. Phys.}

\newcommand*\mnras{Mon.~Not.~Roy.~Astron.~Soc.}

\newcommand*\pasp{PASP}

\newcommand*{\ksM}{\text{km/s Mpc$^{-1} $}}
\usepackage{soul}
\setulcolor{blue}
\setstcolor{red}
\sethlcolor{green}

\def\be{\begin{equation}}
\def\ee{\end{equation}}
\def\bea{\begin{eqnarray}}
\def\eea{\end{eqnarray}}

\def\ptl{\partial}

\begin{document}

\title{Reference level of the vacuum energy density of the Universe and astrophysical data}

\author{Balakrishna S. Haridasu}
\email{haridasu@roma2.infn.it}
\affiliation{Dipartimento di Fisica, Universit\'a di Roma "Tor
Vergata"
}%
\affiliation{Sezione INFN, Universit\'a di Roma "Tor Vergata", Via della Ricerca Scientifica 1, I-00133, Roma, Italy}

\author{S.L. Cherkas}
\email{cherkas@inp.bsu.by}
\affiliation{Institute for Nuclear Problems, Bobruiskaya
11, Minsk 220030, Belarus}
\author{V.L. Kalashnikov}
\email{vladimir.kalashnikov@uniroma1.it }
\affiliation{Facolt\'a di Ingegneria dell'Informazione, Informatica e Statistica, Sapienza Universit\'a di Roma, Via Eudossiana 18
00189 - Roma, RM, Italia}
\affiliation{Institute of Photonics, Vienna University of Technology, Gusshausstrasse 27/387,
            Vienna A-1040, Austria}
            
\date{\today}

\begin{abstract}
  An extended framework of gravity, in which the first Friedmann equation is satisfied up to some constant due to violation of gauge invariance, is tested against astrophysical data: Supernovae Type-Ia, Cosmic Chronometers, and Gamma-ray bursts. A generalized expression for the Friedmann equation, including the possible vacuum contributions, is suggested, and two particular cosmological models with two independent parameters are considered within this framework and compared on the basis of the likelihood analysis. One of the models considered includes contribution of the residual vacuum fluctuations to the energy density and places the limit on the UV cutoff scale as $k_{max} = 12.43^{+0.9}_{-1.6} [M_p/\sqrt{2+N_{sc}}]$, where $N_{sc}$ is the number of minimally coupled scalar fields. Model comparison using the Akaike information criteria and  Bayesian evidence shows a preference for the conventional $\Lambda$CDM over the extended models. A more general model with three parameters is considered within which an anti-correlated behavior between the dynamical vacuum fluctuations contribution and a negative cosmological constant was found. The result is an upper limit of $\Omega_{\Lambda} \lesssim -0.14$ at $95\%$ C.L., which is only mildly disfavored ($\ln\mathcal{B} = -1.8$) with respect to $\Lambda$CDM.
   \end{abstract}

\maketitle


\section{Introduction}
\label{sec:Introduction}
It is well-known that the reference level of energy density in the Minkowski space-time could be arbitrarily chosen as it is not immediately related to the observable data \cite{Landau1982}. Thus, there is no immediate problem with the appearance of large values of the mean energy density of a vacuum state. Formally, the vacuum energy density could be nullified by some renormalization procedure \cite{Collins86}, such as the Pauli-Villars \cite{Slavnov1977}, dimensional \cite{Breitenlohner1977,akhmedov2002vacuum}, or point-splitting regularizations \cite{Birrell82}. This situation is more acute in the general relativity (GR) because, though the space-time in GR locally looks like the Minkowski one, the uniform energy density and pressure have to result in the expansion of the universe. The vacuum energy density calculated with the UV cut-off at a Planck level, would lead to a very high expansion rate of the universe in contradiction to the astrophysical observations \cite{Weinberg89}.

Several approaches were proposed to avoid this issue, among which the first is to assume that a significant value of the vacuum energy does not exist in reality, implying that the renormalization procedure is not a ``technical trick" but has a physical meaning. However, investigations of numerous quantum mechanical systems, such as two-atom molecules \cite{Landau1981}, or atomic nuclei \cite{Eisenberg1972} insist on a reality of the ground state energy of the quantum oscillator. One could
not say that field oscillators differ principally from the
others. Besides, an expectation of a ``good" quantum theory suggests the absence of infinities, as in the case of the string theory \cite{Polchinski1998}. Let us imagine that the (super)string theory \cite{Kaku2012} will be able to produce masses of all particles and all the fundamental interactions after successful compactification from 10 to 4 dimensions. In that case, there is no renormalization required due to the extended nature of a string. For an ordinary field theory, it will be seen as a number of ``sum rules" that have to be satisfied \cite{Visser18}.

That leads to an alternate possibility to assume that the vacuum energy density indeed exists, but the contributions from different fields compensate each other mutually. The example is the exact supersymmetry in which the number of bosonic and fermionic degrees of freedom are equal. Additionally, masses of particles and their super-partners are also identical. At present, no supersymmetric particles have been found \cite{Autermann2016}. Nevertheless, an idea of mutual compensation is not directly related or specific to the supersymmetry and has been first suggested far before the supersymmetry conjecture \cite{Pauli1971}. If the numbers of fermionic and bosonic degrees of freedom are different, then there is no compensation of the main part of the vacuum energy density $\rho_{v}\sim M_p^4$, where $M_p=\sqrt{\frac{3}{4\pi G }}$ is the reduced Planck mass. Thus, one needs an alternate explanation for the fact that the main part of the vacuum energy does not contribute to gravity \cite{Padmanabhan2014, Josset2017, Percacci2018}.

Requiring the nongravitating vacuum energy together with the fact that there is no invariant vacuum state infers the hints for a modification of the theory of gravity. A vacuum energy problem, in this sense, is an excellent gift for theoreticians. It points to the direction of a GR modification, namely, to the violation of the gauge invariance, which allows choosing an arbitrary energy density level in the Friedmann equations. A version of such a theory was proposed in \cite{Vesti}, where the Friedmann equation is satisfied up to some (arbitrary) constant and takes the form
\be
-\frac{1}{2}M_p^2\left(a^{\prime 2}+{ \mathcal K } a^2\right)+\rho
a^4 = const\equiv-\wp,
\label{fried}
\ee
where the prime denotes the derivative of the scale factor $a$ w.r.t the conformal time, and the energy density $\rho$ includes all kinds of matter. Like in the GR, three types of uniform spacial curvatures $\mathcal K=\{-1,0,1\}$ are possible, but they are not related to the critical density.

One has to note that \Cref{fried} has been deduced in some particular gauge \cite{Vesti}
\be
ds=a^2(d\eta^2-e^{4\lambda}dr^2-e^{-2\lambda}r^2(\sin^2\theta d\phi^2+d\theta^2)),
 \ee
where the function $\lambda(r/\mathcal R_0)$ is related to the comoving distance $\chi$  as $r\,
e^{-\lambda(r/\mathcal R_0)}=\mathcal R_0\Phi_{\mathcal
K}\left({\chi}/{\mathcal R_0}\right)$,
\begin{equation}
\label{eqn:luDk}
\Phi_{\mathcal K}(x) \equiv \left\{
\begin{aligned}
& \sin(x),  &&\text{for}\,\,\mathcal K=+1 \\
& x,  &&\text{for}\,\,\mathcal K
=0  \\
& \sinh(x),   &&\text{for}\,\,\mathcal K=-1.
\end{aligned}\right.
\end{equation}
and $6\mathcal R_0^{-2}$ is the present spatial curvature of the universe.

Using this preferred gauge implies a violation of the gauge invariance. As will be shown below, the main part of the vacuum energy density scales as $a^{-4}$ in this case, and the constant on the right-hand side of \Cref{fried} compensates it. Then, we converge to the GR if the main part of the vacuum energy is compensated exactly, but some residual value could remain. Here we will assume that the constant $\wp$ on the right-hand side of \Cref{fried} is this residual value and, at the same time, $\rho$ on the left-hand side of \Cref{fried} is an energy density without the main part of vacuum contribution. Such residual $\wp$ is equivalent to some amount of ``invisible'' (i.e., unperturbable) radiation, which can be either positive or negative. From high quality observational astrophysical data, one could determine the effective equation of state for all content of the universes and the constant  $\wp$, which has to take some unique value for our particular universe.

The ``high-redshift'' observations of cosmic microwave background radiation (CMB) \cite{Hinshaw2013,Planck2016, Planck18_parameters} have a very high constraining power on the cosmological parameters. However, the interpretation of CMB and baryon acoustic oscillations (BAO) \cite{Eisenstein05} data stays, at least mildly, model-dependent, i.e., they are based on the standard GR framework. The interpretation of these data in the current extended context (\ref{fried}) is not a trivial problem, which we intend for a future investigation.  

The model-independent low-redshift supernovae (SN) dataset gives the rigorous constraints on the cosmological models, as well. The SN datasets have undergone substantial improvements in the last decade, with more SN and more robust statistical methods to compile a homogeneous dataset that can be implemented to test the cosmological models. The most used compilation provides with the joint light-curve analysis (JLA) dataset in Betoule et al. \cite{Betoule14}, with 740 selected SN up to a redshift of $z \sim1.4$, which has been recently updated to $\sim 1050$ SN in \cite{Scolnic17}. Gamma-ray bursts (GRB) are the most energetic explosions in the universe. They are detectable up to very high redshifts $(z \sim 8)$. Therefore, they can be used to study the universe expansion rate when the empirical correlations between the spectral and intensity properties are appropriately calibrated. We consider the GRB dataset comprising of 109 observations compiled in \cite{Wei10}, which uses the well known Amati relation \cite{Amati08}. 

These two datasets are complemented with the cosmic chronometers (CC) \cite{Jimenez01}, which provide the measurements of the expansion rate at the different redshifts and are the useful observable for estimating the Hubble constant value-$H_0$ \cite{Moresco12, Yu17, Lukovic18, Gomez-Valent18, Haridasu18_GP}.

However, the quality of the aforementioned observational data is not yet sufficient to derive an equation of state for the overall energy density $\rho$ in \Cref{fried}. Thus, some concrete/simplified models with a specified vacuum nature are required, and here we consider two particular models. The first $\overline{\Lambda{\rm CDM}}$ model is a simple extension of the well-known ${\Lambda{\rm CDM}}$ on basis of \Cref{fried}. The second one is the vacuum fluctuations 
domination model (VFD) \cite{conf} in which the residual vacuum energy density and pressure arise due to universe expansion (elaborated in \Cref{sec:Theory}).

The organization of the paper is as follows: In \Cref{sec:Theory} we describe the theoretical framework and cosmological observables in \Cref{sec:Data} we then briefly present the datasets and model-selection criteria implemented. In \Cref{sec:Results} we present our results for cosmological inferences, constraints on parameters and finally summarize our conclusions in \Cref{sec:Conclusions}.

\section{Residual vacuum energy density and pressure}
\label{sec:Theory}
A scalar field $\phi(x)$ is a convenient and straightforward candidate to study the vacuum energy density problem, for which the energy-momentum tensor could be written in the form \cite{Birrell82}
\be
T_\mu^\nu=\frac{1}{2}\left(\ptl_\mu\,\phi\ptl^\nu\phi+\delta_\mu^\nu
m^2\phi^2\right).
\label{t}
\ee

After quantization of the scalar field, the expectation mean value over a vacuum state of \Cref{t} can be written in the hydrodynamic form for some resting (i.e., having 4-velocity $u^\mu=\{1,0,0,0\}$ ) medium:
\be
\langle 0|T_{\mu}^\nu|0 \rangle =(\rho_v+p_v)\delta_\mu^0\delta^\nu_0-p_v\delta_\mu^\nu.
\label{1}
\ee

The vacuum substance permitted by a covariance must satisfy $p_{v}=-\rho_{v}$, however, the quantity $\langle 0|T_{\mu}^\nu|0 \rangle$ is not a perfect tensor for two reasons: the non-invariance of the vacuum state $|0\rangle$ relative to general coordinate transformations, and the non-covariant character of the UV cut-off.

Thus, the vacuum energy includes the  main part of  $\rho_{v}\sim M_p^4$ as well as other contributions produced by:
\par\noindent
i) the masses of particles and condensates, which depend on the particular mechanism of the particle mass generation,
\par\noindent
ii) the curvature of a space-time, which depends on the universe expansion rate.

For the type of i), which exists equivalently in the Minkowski space-time, one has \cite{Cherkas07}:
\begin{equation}
    \begin{split}
        \rho_v &= \frac{1}{4 \pi^2a^4}\int_0^{k_{max}}k^2\sqrt{k^2+a^2m^2} dk \\
 &\approx \frac{1}{16\pi^2}\biggl(\frac{k_{max}^4}{a^4}+\frac{ m^2 k_{max}^2}{a^2} \\ 
 &\qquad \qquad + \frac{m^4}{8}\left[1+2\ln\left(\frac{m^2a^2}{4k_{max}^2}\right)\right]\biggr),~
    \end{split}
    \label{mass}
\end{equation}
where $k_{max}\sim M_p$ as it was assumed in \cite{Cherkas07}. The term of $\rho_{v}\sim k_{max}^4 a^{-4}$ scales with $a$ as an invisible radiation and is compensated by an arbitrary constant in \Cref{fried} up to its residual value 
\be
\label{eqn:Invrad}
\wp\equiv\frac{1}{2}M_p^2H_0^2\,\Omega_i,
\ee
where $H_0$ is the Hubble constant and a dimensionless quantity $\Omega_i$
has been introduced. 

The next term in \Cref{mass} corresponds to a substance with the equation of state $p_v=-\frac{1}{3}\rho_v$, which was discussed widely \cite{John96,Melia15}, but was considered as an overall (total) equation of state, without the discussion of its origin. Fermions give the contribution of the opposite sign, and according to the observations (i.e., absence of the fast universe expansion), it is expected \cite{Visser18} that the mutual contributions of fermions and bosons compensate each other in every massive term to the accuracy of the order of the critical density. The last term $\rho_{v}\sim m^4$ corresponds roughly to the cosmological constant \cite{Zeldovich67}, with the logarithmic accuracy. One could assume that this term is almost compensated by taking into account the contribution of the condensates, which are of the same order of $\sim m^4$.

The energy density of the ii)-type is expressed through the time derivatives of the universe scale factor \cite{conf,Cherkas07}:
\begin{align}
    \label{eqn:rhoandp}
\rho_v&=\frac{a^{\prime 2}}{2a^6}M_p^2S_0,
\end{align}

 The corresponding pressure is
\begin{align}
    \label{eqn:rhoandp1}
p_v&=\frac{M_p^2S_0}{a^6}\left(\frac{1}{2}a^{\prime
2}-\frac{1}{3}a^{\prime\prime}a\right),
\end{align}
where,
\begin{equation}
    \label{eqn:s0}
    S_0 = \frac{k_{max}^2}{8 \pi^2 M_p^2}
\end{equation}
is determined by the ultra-violet (UV) cut-off of the comoving momenta. Accounting for additional scalar fields, the quantity $k_{max}^2$ in \Cref{eqn:s0} should be replaced by $k_{max}^2\rightarrow (N_{sc}+2)k_{max}^2$, which takes into account two degrees of freedom corresponding to the gravitational waves \cite{Cherkas07}. In a minimal variant of the Standard Model, there exists the SU(2) duplet of the complex scalar fields \cite{Commins1983}, i.e., four scalar degrees of freedom $N_{sc}=4$ before the spontaneous symmetry violation (so-called ``electroweak transition''). At present, after the symmetry violation, only a single Higgs field contributes, but Pauli's sum rules insist on the existence of some new bosons \cite{Visser18}.

The above definitions follow the continuity equation
\begin{equation}
    \rho_v^\prime+3\frac{a^\prime}{a}(\rho_v+p_v)=0.
    \label{3}
\end{equation}
Substituting the vacuum contributions of \Cref{mass,eqn:rhoandp} with adding a pressureless matter ($\Omega_m$) into \Cref{fried} result in a general cosmological model:
\begin{widetext}
\begin{equation}\label{eq:Hubble}
  E(a)^2 =  \frac{(1- S_0-\Omega_{m}-\Omega_{\Lambda}
  -\Omega_k ){a^{-4}} + {\Omega_{m}}{a^{-3}}+{\Omega_{k}}{a^{-2}} +
   \Omega_{\Lambda}+\Omega_L \ln a}{1 - S_0 a^{-2} },
\end{equation}
\end{widetext}
where $E(a) =H(a)/H_0$.
From \Cref{fried}, \Cref{eq:Hubble} holds for an arbitrary signature of the space metric of \Cref{eqn:luDk}. The ``curvature'' term ${\Omega_{k}}{a^{-2}}$ originates from both real spatial curvature and vacuum contribution, which looks like a fluid with the equation of state $p_v=-\frac{1}{3}\rho_v$. The total amount of radiation is $1- S_0-\Omega_{m}-\Omega_{\Lambda}-\Omega_k$. One has to note that an ``invisible'' radiation given by  (\ref{eqn:Invrad}) is indistinguishable from the real radiation. However, the real radiation contribution is rather low at low redshifts and does not contribute to the dynamics of the expansion history.

Although a number of the contributors are possible in \Cref{eq:Hubble} due to violation of the gauge invariance, we will consider below only two particular models with two independent parameters and, besides assuming a flat universe $\mathcal K=0$ for simplicity. The first $\overline{\Lambda\mbox{CDM}}$-model corresponds to $\Omega_k=0$, $S_0=0$, $\Omega_L=0$ in \Cref{eq:Hubble}.  It has the independent parameters $\Omega_m$ and $\Omega_\Lambda$. In contrast to the well-known $\Lambda\mbox{CDM}$, the matter density and the cosmological constant are not constrained by the relation $\Omega_\Lambda+\Omega_m=1$ by virtue of \Cref{fried}. The second model is the vacuum fluctuations-dominant model ($\mbox{VFD}$), corresponding to $\Omega_k=0$, $\Omega_\Lambda=0$, $\Omega_L=0$ with the non-zero parameters $S_0$ and $\Omega_m$. 

The denominator in \Cref{eq:Hubble} urges $S_0>1$, otherwise, a Big Rip would have occurred by the current epoch. The VFD implies that among all the substances filling the universe, the residual energy density and pressure of a vacuum are dominant.  In regards of dark energy and dark matter, we possibly do not require these exotic substances when a ``correct'' vacuum contribution is taken into account. The appropriate consideration of the vacuum energy could be an alternative to the quintessence models, which conform with various datasets better or are at least statistically equivalent to $\Lambda$CDM \cite{Sola17a, Sola17b, Park18a, Park18b, Tosone18}. 

While the EoS of VFD model \cite{Cherkas18}, in some sense, looks similar to that of \cite{Akarsu14}, where the hybrid expansion law was implemented, these models, behave differently at higher redshifts due to different gravity theory frameworks. The VFD-model behaves asymptotically as the Milne-like one, whereas the model of \cite{Akarsu14} demonstrates a deceleration, although  without a matter-dominated regime. Finally, we also test a further extension of the VFD model by allowing for the $\Omega_{\Lambda}\neq 0$ component (hereafter V$\Lambda$CDM) appealing to assess the nature of a cosmological constant, which could originate from both vacuum fluctuations and false vacuum condensates.

\section{Dataset and {Methods}}
\label{sec:Data}

\begin{figure*}[!ht]
    \centering
    \includegraphics[scale=0.43]{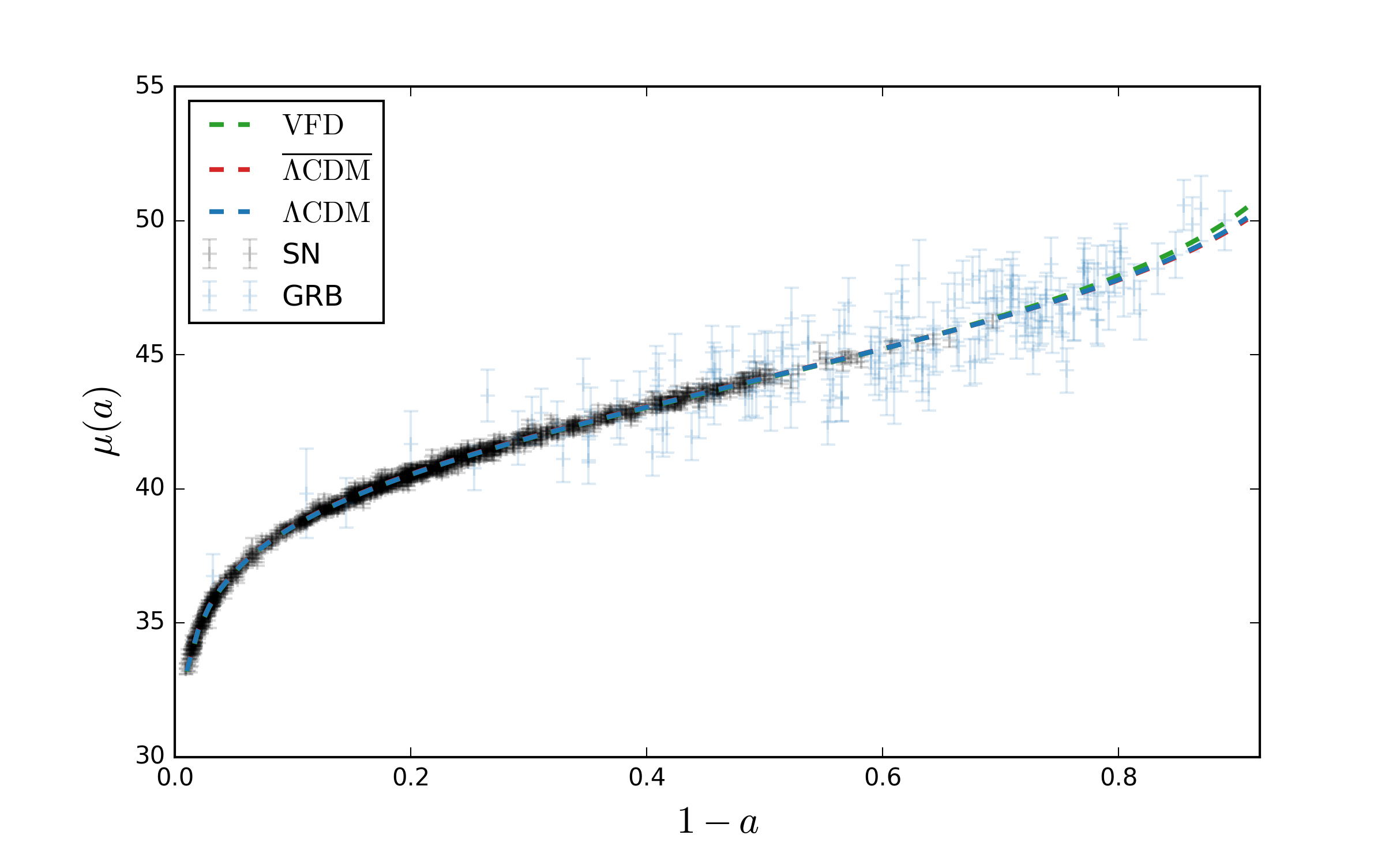}
    \includegraphics[scale=0.43]{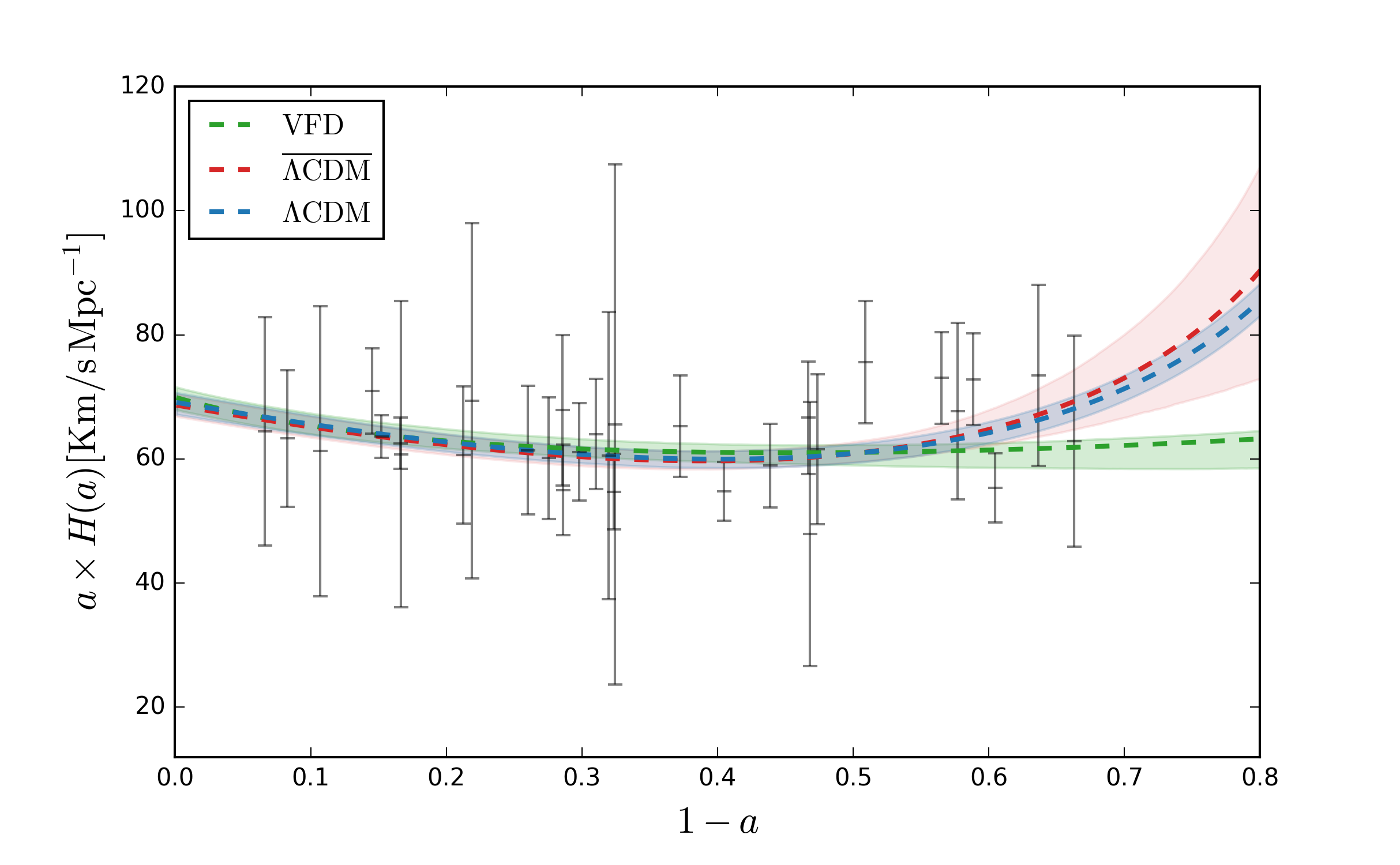}
    \vspace{-1cm}
    \hspace{2cm}
    \caption{\textit{Left Panel:} The distance modulus of the SN and GRB dataset plotted against the best-fit models. \textit{Right Panel:}  The model-dependent expansion histories obtained using the SN+CC+GRB datasets, best-fit (dashed), and the dispersion between the $16^{\rm th}$ and $84^{\rm th}$ percentiles are shown as a shaded region.}
    \label{fig:VLCDM_CC+GRB+SN_h}
\end{figure*}

We utilize a combination of the astrophysical datasets under an essential requirement of their independence on any cosmological assumptions. A summary of such datasets is presented here:

\textit{Supernovae ${\rm (SN)}$:} The Pantheon compilation of $\sim 1050$ SNe observations presented in \cite{Scolnic17} has improved the statistical precision and the highest redshift ($z\sim 2$) to which the distances have been measured. We take advantage of this dataset, which already remains a mild improvement over the previous one \cite{Betoule14}. The later dataset eases the analysis in comparison to the earlier one by marginalizing the supernovae standardization parameters a priori in a model-independent way.

\textit{Cosmic Chronometers ${\rm (CC)}$:} Differential dating of galaxies was proposed as the means to estimate the Hubble constant in a model-independent way \citep{Jimenez01}. These estimations are related to the synthetic spectra of simple stellar populations and the models of stellar evolution. Recently a very robust characterization of the differential aging has been tested \citep{Moresco16}, and it provides an estimation of $H(z)$ at $6\%$ accuracy (see also \cite{Moresco18} for the recent review on the framework of differential dating). In this work, we adopted the measurements provided by \cite{Simon05, Moresco12, Moresco15, Moresco16, Ratsimbazafy17} listed in Table $2$ of \cite{Haridasu18_GP}, which comprises $31$ measurements of $H(z)$ over the redshift interval $z \in (0.0798,1.965)$. One has to note that the CC dataset prevents from an intrinsic systematic effect due to the assumption of the stellar evolution models, whose effect on the estimation of $H_0$, as it was evaluated in \cite{Haridasu18_GP}, causes an additional systematic error of $\sigma_{sys} \sim 2.5\, \ksM$.

\textit{Gamma ray bursts ${\rm (GRB)}$:} GRBs are observed in a wide range of spectroscopic and photometric redshifts, up to $z \sim 8$, and can be used to probe the high-$z$ universe. As GRBs are not standard candles, they are standardized by utilizing the SN distance modulus in the overlapping redshift range and can provide insights into a higher-redshift evolution of the cosmological models. We implement the GBR likelihood, as was earlier utilized in \cite{Haridasu17} (please refer for further details of likelihood construction). The current dataset comprising of 109 GRBs has been compiled in \cite{Wei10} utilizing the Amati relation \cite{Amati08}. The dataset has 50 GRBs at $z < 1.4$ and 59 GRBs at $z > 1.4$ in a total range of $0.1 < z < 8.1$ \cite{Wei10}.

A simple joint likelihood of these datasets is constructed as 
\begin{equation}
    \mathcal{L}({\bf y}| \Theta) = \mathcal{L_{\rm SN}}\times\mathcal{L_{\rm CC}}\times \mathcal{L_{\rm GRB}},
\end{equation}
which is utilized to perform a Bayesian analysis through MCMC sampling. For this purpose, we use the \texttt{emcee}\footnote{\href{http://dfm.io/emcee/current/}{http://dfm.io/emcee/current/}} \citep{Foreman-Mackey13} package, which implements an affine invariant Metropolis-Hastings sampler. We also utilize the \texttt{getdist}\footnote{\href{https://getdist.readthedocs.io/}{https://getdist.readthedocs.io/}} package to analyze the chains and obtain posteriors. A complete Bayesian analysis has been performed here because a simple Frequentist approach can be insufficient for the needs of the strongly non-Gaussian posterior, which the VFD model predicts. We implement flat/uniform priors on the parameters, which \Cref{Tab:priors} summarizes.

{\renewcommand{\arraystretch}{1.5}%
    \centering
    \begin{table}[ht!]
        \caption{Priors implemented in our Bayesian analysis, as relevant for the respective models.}
        \label{Tab:priors}
        \begin{tabular}{ccccc}      
            \hline
            \hline
            parameter  & VFD &  {$\overline{\Lambda {\rm CDM}}$} & $\Lambda$CDM & V$\Lambda$CDM \\ 
            \hline
            $H_0$    & [50.0,100.0] & [50.0,100.0] & [50.0,100.0] & [50.0,100.0] \\
            $\Omega_{m}$  & [0.0,3.0] &  [0.0,1.0] &  [0.0,1.0] &  [0.0,3.0] \\ 
            $S_0$    & [1.0,10.0] &   0.0 & 0.0 & [1.0,5.0] \\
            $\Omega_{\Lambda}$  & 0.0 &  [0.0,1.0] & - &  [-3.0,1.0] \\
            \hline
            \hline
        \end{tabular}
    \end{table}
}

It is often convenient to take the conventional $\Lambda\mbox{CDM}$ as a reference model for comparison and establish corresponding statistical criteria. We implement the widely used Akaike information criteria (AIC) \cite{Akaike74} for model selection \cite{Burnham98}. AICc, corrected for a number of data points to the second-order is written as
\begin{align}
    \rm{AICc} &= -2\log{\cal L}^{max} + 2 N_p + \frac{2N_p(N_p+1)}{N_d-N_p-1},
\end{align}

\noindent where $N_p$ is a number of parameters and $N_d$, is the number of data points. The model preference is estimated by evaluating $\Delta$AICc, as a difference in the AICc value of the model in comparison to the reference model ($\Lambda$CDM). A positive value of $\Delta$AICc indicates that the reference model is preferred over the model in comparison\footnote{See \cite{Trotta08, Trotta17} for more details.}. 

Alongside AIC, we also compute the Bayesian evidence \cite{Trotta17, Heavens17, Haridasu18_GP}, owing to the ease of implementation with our MCMC analysis \cite{Heavens17a}\footnote{We utilized the MCEvidence package publicly available at \href{https://github.com/yabebalFantaye/MCEvidence}{https://github.com/yabebalFantaye/MCEvidence}.}. Given the observations $\mathcal{D}$ and $\mathcal{M}(\Theta)$ of the assumed model, find the posterior distribution w.r.t. $\Theta$ according to the Bayes' rule

\begin{equation}
    \label{eqn:BayesRule}
    p(\Theta|\mathcal{D},\mathcal{M}) = \dfrac{p(\mathcal{D}|\Theta,\mathcal{M})\pi(\Theta|\mathcal{M})}{p(\mathcal{D}|\mathcal{M})},
\end{equation}
where $\pi(\Theta|\mathcal{M})$ is the prior and $p(\mathcal{D}|\mathcal{M})$ is the Bayesian `evidence' and can be written as
\begin{equation}
    p(\mathcal{D}|\mathcal{M}) = \int p(\mathcal{D}|\Theta,\mathcal{M})\pi(\Theta|\mathcal{M}) \textrm{d}\Theta.
\end{equation}

A more complex model, usually with higher likelihood and more parameters is penalized with the larger prior volume. Comparing the `evidence' for two given different models $\mathcal{M}_A$ and $\mathcal{M}_B$, with priors $\pi(\mathcal{M}_A)\, , \pi(\mathcal{M}_B$), can be inferred using the Bayes factor as
\begin{equation}
    \frac{p(\mathcal{M}_A|\mathcal{D})}{p(\mathcal{M}_B|\mathcal{D})} =\frac{\pi(\mathcal{M}_A)\times p(\mathcal{D}|\mathcal{M}_A)}{\pi(\mathcal{M}_B)\times p(\mathcal{D}|\mathcal{M}_B)},
\end{equation}
where $p(\mathcal{M}_i|\mathcal{D})$ is the posterior and the ratio of the evidences on the left hand side is called the Bayes factor ($\mathcal{B}$). We compute the $ \ln\mathcal{B}$, which a negative value indicates that the reference model is favored over the model in comparison. \Cref{Tab:CC+GRB+SN} presents the results w.r.t the favored model. The strength of the preference/rejection of a model is often gauged in terms of Jeffrey's scale \cite{Kass95} as was discussed and implemented in earlier works \cite{Nesseris13,Camarena18,Romero18}.


\section{Results}
\label{sec:Results}

{\renewcommand{\arraystretch}{1.9}%
    \centering

    \begin{table*}[ht!]
    \caption{$68\%$ C.L. constraints for the VFD, $\overline{\Lambda\mbox{CDM}}$, $\Lambda$CDM and V$\Lambda$CDM models obtained using the SN+CC+GRB datasets. In evaluating the information criteria presented in the last row, $\Lambda$CDM is taken as the reference model. Here $^{*}$ indicates a derived quantity, and we report the $95\%$ C.L. limits for unconstrained parameters. Here $H_0$ is expressed in the units of $\ksM$.}
        \label{Tab:CC+GRB+SN}
        \begin{tabular}{cccccccc}
            \hline
            \hline
            parameter  & \multicolumn{2}{c}{VFD} &  \multicolumn{2}{c}{$\overline{\Lambda\mbox{CDM}}$} &  \multicolumn{2}{c}{$\Lambda$CDM} & {V$\Lambda$CDM} \\
            \hline
             & b.f  & $1\sigma$ &b.f  & $1\sigma$ &b.f  & $1\sigma$ & $1\sigma$\\
            \hline
            $H_0$   & $69.8$  & $69.8\pm 1.9        $ & $68.9$ &$68.7\pm 1.8       $ & $68.9$ & $68.9\pm 1.8        $ & $69.6\pm 1.8               $\\
            $S_0$   & $1.60$  &  $1.98^{+0.24}_{-0.52}      $ & $0$ & --  & $0$ & --  & $> 2.16                    $\\
            $\Omega_{m}$ & $0.87$  & $0.64^{+0.34}_{-0.19}      $  & $0.29$& $0.276^{+0.090}_{-0.074}   $ & $0.302$& $0.303\pm 0.020   $ & $> 0.594                   $ \\
            $\Omega_{\Lambda}$ & $0$ & -- & $0.703$ & $0.708\pm 0.041            $ & ${0.698}^{*}$ & ${0.697\pm 0.020}^{*}$ & $< -0.14                    $  \\
            $\Omega_{i}^{*}$ & $-1.48$ & $-1.62^{+0.22}_{-0.09}    $ & $0.007$ & $0.016^{+0.037}_{-0.048}   $ & \multicolumn{2}{c}{$0$ } & $<-1.90$  \\
            \hline
            $\chi^2_{\rm b.f}$ & \multicolumn{2}{c}{$1215.34$ } &  \multicolumn{2}{c}{$1213.24$} &  \multicolumn{2}{c}{$1213.26$ } &  1214.16 \\
            $\Delta {\rm AICc}$ &  \multicolumn{2}{c}{$4.02$ }  &  \multicolumn{2}{c}{$1.92$ }  &  \multicolumn{2}{c}{$0$ } &  {$4.86$ }\\
            $ {\ln{\mathcal B}}$ &  \multicolumn{2}{c}{$-8.0$ }  &  \multicolumn{2}{c}{$-2.6$ }  &  \multicolumn{2}{c}{$0$ } &  {$-1.9$ }\\
\hline
            \hline
        \end{tabular}
   \end{table*}
}

\begin{figure}[!ht]
    \includegraphics[scale=0.34]{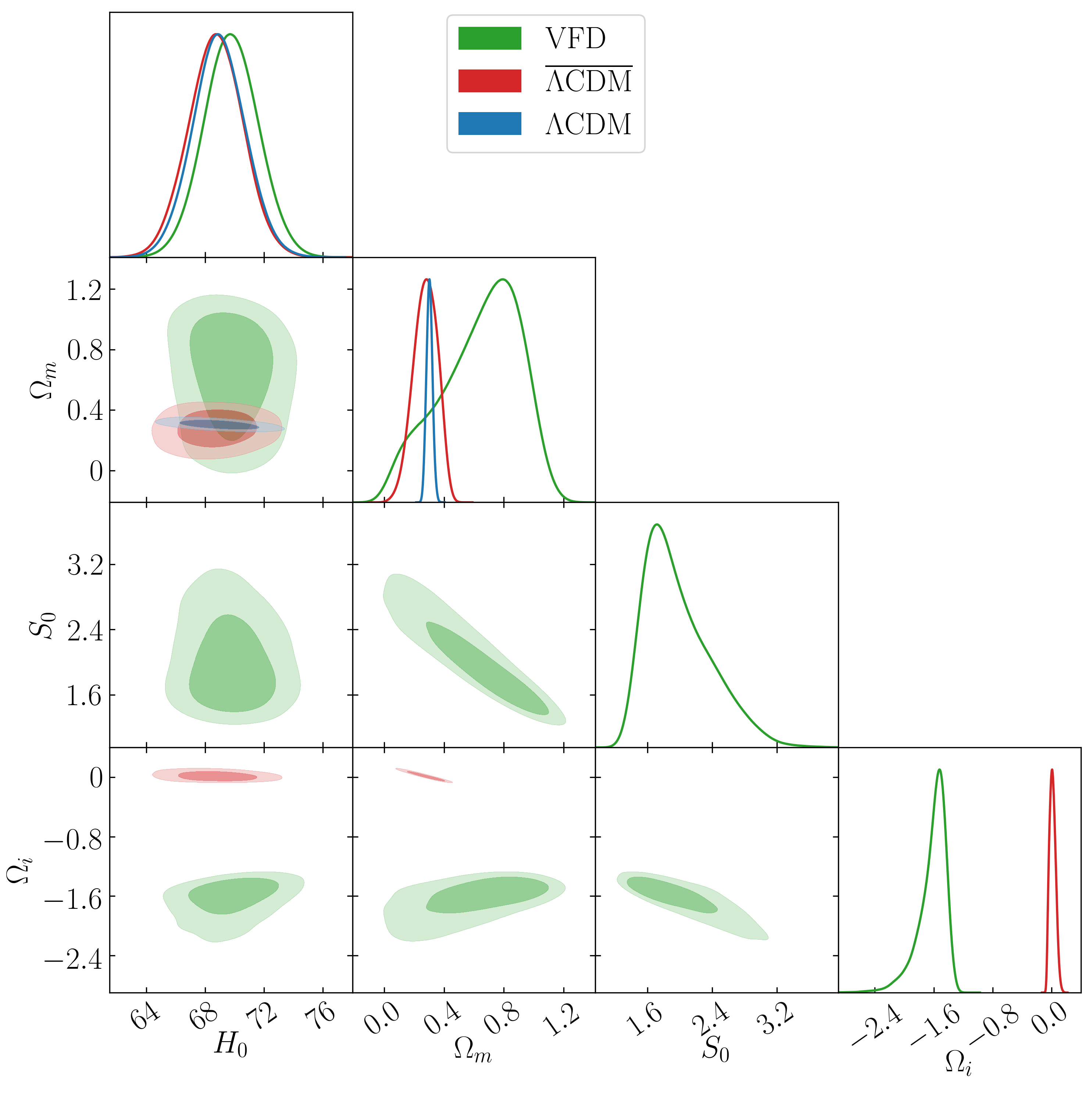}
    \caption{Confidence regions for the VFD, $\overline{\Lambda{\rm CDM}}$ and $\Lambda$CDM models
    obtained using the CC+SN+GRB dataset, corresponding to the constraints presented in \Cref{Tab:CC+GRB+SN}. }
    \label{fig:VCDM_CC+GRB+SN}
\end{figure}

As shown in the left panel of \Cref{fig:VLCDM_CC+GRB+SN_h},  all the models seem almost identical, while showing variation for $z \gtrsim 1.5$.
In \Cref{Tab:CC+GRB+SN}, we present the $1\sigma$ constraints obtained from our joint analysis. As can be clearly seen, also from the confidence regions shown in \Cref{fig:VCDM_CC+GRB+SN}, $\Omega_m$ is far less constrained in the VFD model in comparison to $\Lambda$CDM and $\overline{\Lambda{\rm CDM}}$, allowing for $\Omega_m \sim 1$. That is due to the strong anti-correlation between the $\Omega_m$ and $S_0$ parameters, which provides the upper limit on $\Omega_m$, obeying the lower limit of $S_0>1.0$ prior (see \Cref{sec:Theory}). Given the very large uniform prior parameter space allowed for the $S_0$ parameter, we find that the data is able to constrain it, which however, could benefit from the additional data that can refine the posteriors on $\Omega_m$ (see for example \cite{Singh19}). In fact, $\Omega_m-S_0$ parameter space indicates the improvement of the fit over $S_0 = 0$, which corresponds simply to the Einstein-de Sitter like Universe with $\Omega_m\sim1$ \cite{Einstein32}. Also, following \Cref{eqn:s0}, we can immediately translate the constraint on $S_0$ to obtain the constraint on the UV cut-off scale $k_{max} = 12.43^{+0.9}_{-1.6} [M_p/\sqrt{2+N_{sc}}]$, in the units of the reduced Planck mass.

\begin{figure*}[!ht]
    \centering
    \includegraphics[scale=0.43]{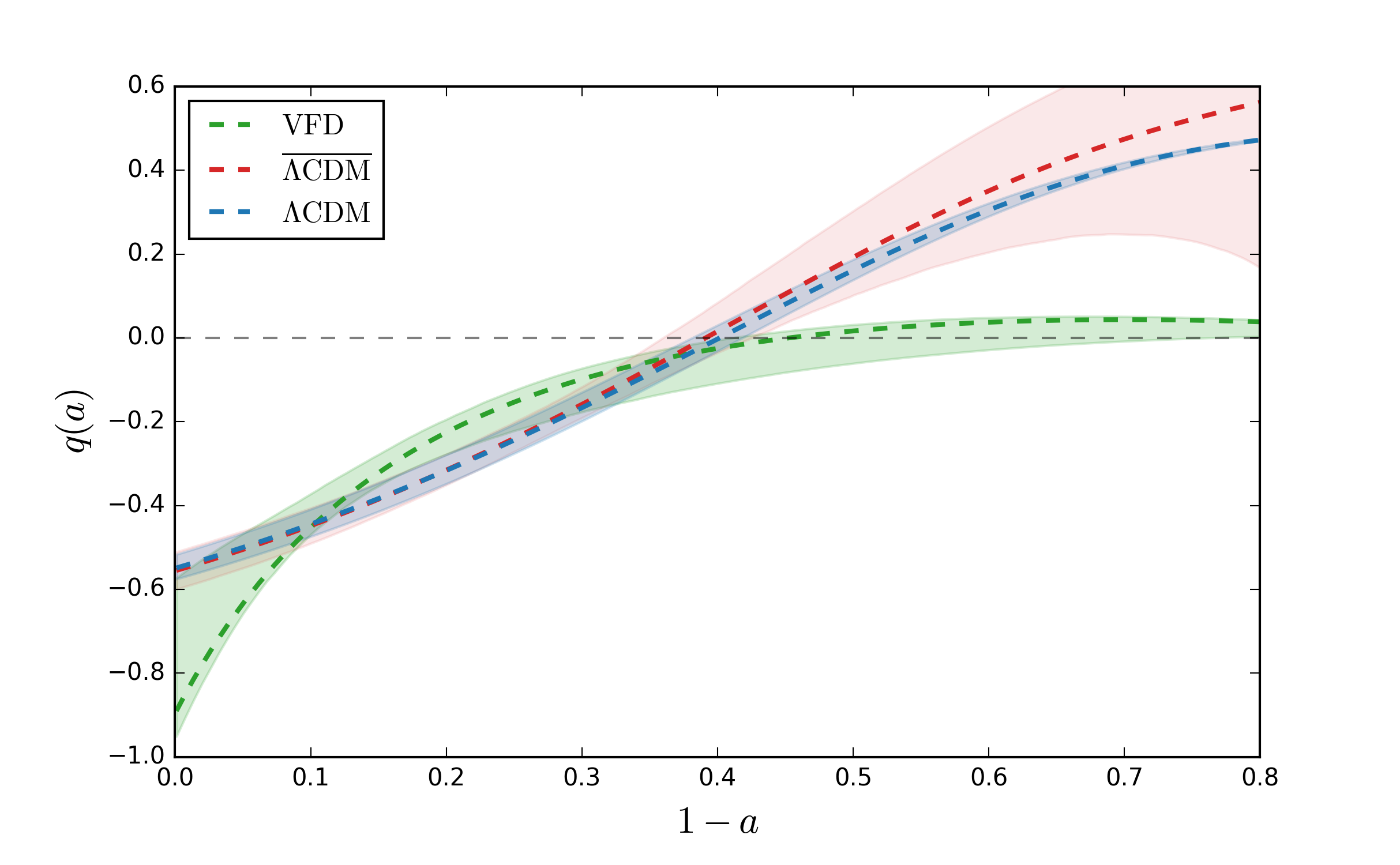}
    \includegraphics[scale=0.43]{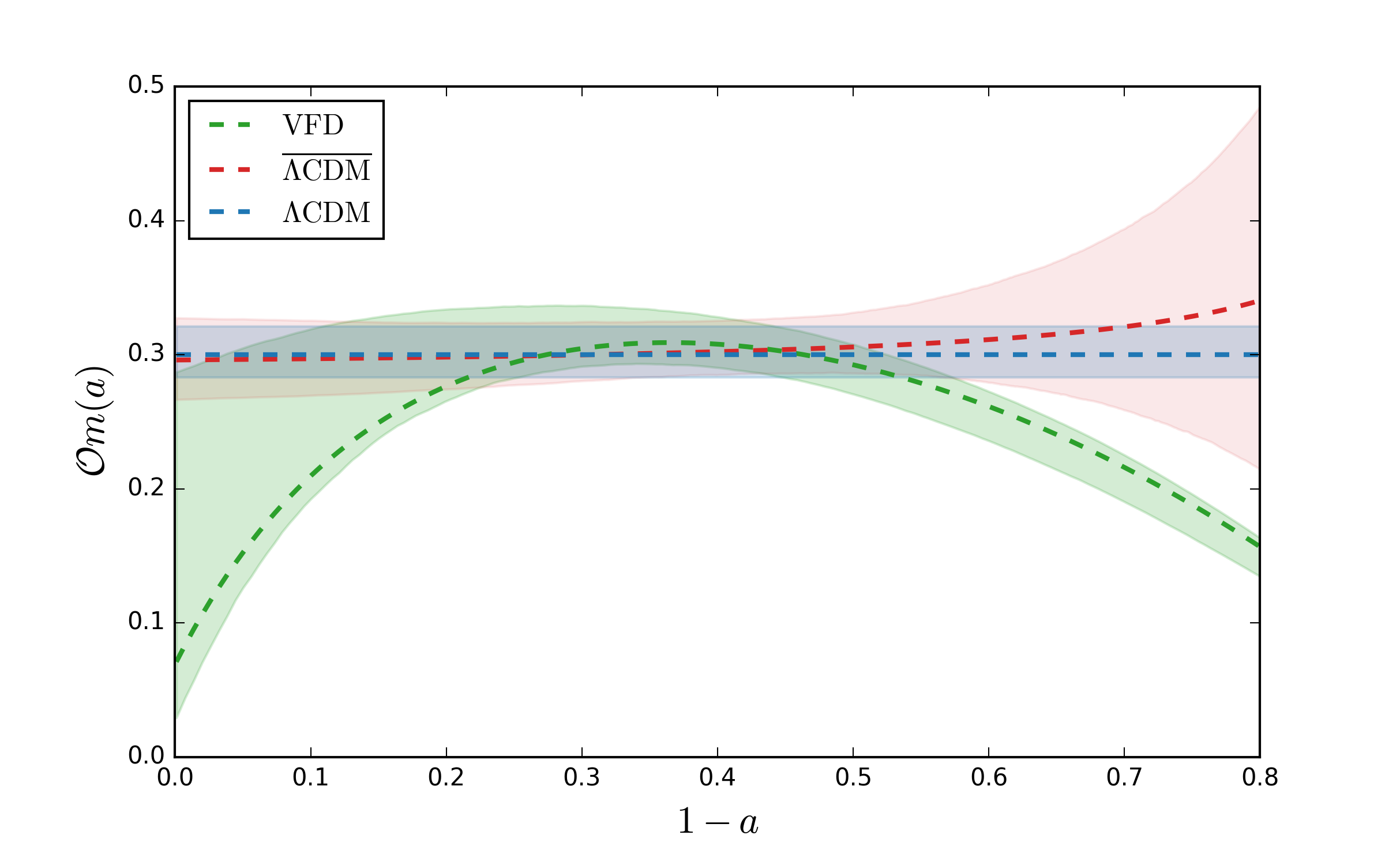}
    \caption{\textit{Left Panel:} Comparison of the model predictions for the deceleration parameter $q(a)$ with SN+CC+GRB data, best-fit (dashed) and the corresponding dispersion between the $16^{\rm th}$ and $84^{\rm th}$ percentiles.
     \textit{Right Panel:}
  "Effective" amount of matter given by  $\mathcal{O}_m(a)$ from \Cref{om}. }
    \label{fig:VLCDM_CC+GRB+SN_q}
\end{figure*}

In our analysis, the constraints on $H_0$ are driven by the CC dataset and are consistent with the earlier reported model-independent estimates in \cite{Haridasu18_GP, Gomez-Valent18}.  While we do not intend any implications in the context of the well-known $H_0$-tension \cite{Bernal16, Lukovic16, Riess19, Wong19}, we notice a small shift towards higher values ($\Delta H_0 \sim 1 $) in the VFD model, which follows the faster acceleration rates, as shown in the left panel of \Cref{fig:VLCDM_CC+GRB+SN_q}. 

The value of the ``invisible radiation'' $\Omega_i$ determined by \Cref{eqn:Invrad} is presented in the last row of \Cref{Tab:CC+GRB+SN} and in \Cref{fig:VCDM_CC+GRB+SN}. It indicates that $\overline{\Lambda{\rm CDM}}$ insists on the GR framework of $\Omega_i=1-\Omega_m-\Omega_\Lambda\approx 0.016$, while VFD is clearly not a GR-based theory having $\Omega_i=1-S_0-\Omega_m\approx-1.48$. The $\Omega_i$ freedom in $\overline{\Lambda{\rm CDM}}$ model aids to a larger uncertainty of $\Omega_m$ in comparison to $\Lambda$CDM, while keeping the low-redshift dynamics indistinguishable. The effects are in fact noticeable only as a larger dispersion at the higher redshifts (see \Cref{fig:VLCDM_CC+GRB+SN_h,fig:VLCDM_CC+GRB+SN_q}). The relatively small dispersion of $\Omega_i$ for $\overline{\Lambda{\rm CDM}}$ leads to a visible effect in the $a H(a)$ dynamics. \Cref{eqn:Invrad} allows us to place the limits $\wp = -0.81^{+0.11}_{-0.05} [M_p^2H_0^2]$ on the arbitrary constant in \Cref{fried}, in the units of $[M_p^2 H_0^2]$.

The VFD deceleration parameter $q(a)$, shown in \Cref{fig:VLCDM_CC+GRB+SN_q} at the face-value appears to be inconsistent with the deceleration-acceleration transition redshift constrained in \cite{Haridasu18_GP} (see also \cite{Zhang16}). Such $q(z)$ prediction indicates that VFD belongs to a class of the so-called Milne-like (coasting-like behavior) models for $z>0.7$ \cite{Milne35,John96,Dev02} which expand uniformly (i.e, $q(z) = 0$). This class of models have recently attracted attention \cite{Benoit14, Melia15, Steinacker18, Singh19, John19}, partially since the data, when utilized separately, are consistent or even prefers a linear coasting evolution over $\Lambda$CDM at the late-times (see, e.g., \cite{Melia17}), apart from the physical motivations \cite{Gehlaut03, Kumar16, Dev00}. However, a joint analysis of these low-redshift data does not support the claim of \cite{Haridasu17, Lin18}, also within a cosmology-independent analysis \cite{Gomez-Valent18, Haridasu18_GP, Gomez-Valent19, Mukherjee19}. 

In this respect, the VFD model provides an interesting possibility with an early linear coasting-like behavior and a late acceleration with $q(a=1) = -0.74^{+0.16}_{-0.22}$. 
More recently in \cite{Singh19}, the authors studied the primordial nucleosynthesis under a linear coasting background and claimed that the observed helium content could be reconciled by a larger baryon density and without a dark matter \footnote{Please, refer to \cite{Lewis13, Melia14, Lewis16, Bengochea16, Cherkas17, Cherkas18} and 
references therein for the extended discussion and contrasting arguments about the early-time (pre-recombination) behavior of the linear coasting-like models. Also, 
please note that here we have bunched an incomplete list of works analyzing a linear-coasting behavior arising due to different physical considerations.}. Yet an additional confirmation of this fact with some alternative nucleosynthesis analysis is crucially desirable.

It is interesting to use the $\mathcal{O}_m$ diagnostic \cite{Sahni14},
\be
\mathcal{O}_m(z)=\frac{E^2(z)-1}{(1+z)^3-1}
\label{om}
\ee
which was developed as a diagnostic for $\Lambda$CDM being equivalent to $\Omega_m$. As one can see from \Cref{fig:VLCDM_CC+GRB+SN_q} (right panel), the relatively small dispersion of $\Omega_i$ leads to a considerable dispersion of $\mathcal{O}_m(z)$ at $z>1$ for $\overline{\Lambda{\rm CDM}}$ compared to the standard $\Lambda$CDM. For the VFD model, this quantity could be interpreted as an ``effective'' matter content, which is, certainly, very distinct from the actual $\Omega_m$ constraints. In comparison, the earlier model-independent reconstruction in \cite{Haridasu18_GP} (Fig. 6 therein) gives the qualitatively similar predictions for $\mathcal{O}_m(z)$ at $z>2$ as the VFD model (i.e., a downward folding of the curve is implied). It should be noted that this feature was interpreted in \cite{Haridasu18_GP} primarily due to the Baryon Acoustic Oscillation observations at $z\sim 2.4$ \cite{Bautista17,Bourboux17}, which are not considered here. However, it is interesting to note that in the standard GR-like phenomenology, such behavior indicates a negative energy density ($E(z)<E(z)_{\Lambda {\rm CDM}}$), which was earlier mentioned in \cite{Mortsell18, Dutta18, wang18, Zhao17_nature}, and corresponds to the recent model-independent predictions utilizing various datasets in \cite{ Haridasu18_GP, Capozziello19, Gomez-Valent19}. 

On the other hand, we notice a similar folding of the curve towards $z\xrightarrow{} {0}$, which is associated with a `Big Rip' \cite{Caldwell03} future with $w_{\rm de} < -1.0$ and a phantom-like behavior in $\mathcal{O}_m$ diagnostic \cite{Sahni08}. That, in turn asserts the interesting aspects of the VFD model, which, being different in formulation from the standard scenarios, can facilitate distinct phenomenological predictions through the modeling of a single parameter $S_0$.
 
In any case, as it is shown in the penultimate row of \Cref{Tab:CC+GRB+SN}, the AICc statistic indicates a preference for the reference $\Lambda$CDM model w.r.t. both $\overline{\Lambda{\rm CDM}}$ and VFD. The comparison between the $\overline{\Lambda{\rm CDM}}$ and VFD models with the equal number of parameters boils down to the difference $\Delta \chi^2 = 2.1$, which is just about the moderate significance having a very different parameter space. While at face value, what might seem mildly discouraging for the VFD model, we emphasize the advantage of VFD being able to produce diverse phenomenology and also having a physical motivation over a more restricted and yet only a phenomenological $\overline{\Lambda{\rm CDM}}$.While the Bayesian evidence also indicates similar behavior, we find that the VFD model is much more strongly disfavored on the Jeffrey scale with an $\ln\mathcal{B} = -8.0$, which is partly due to the assumed larger prior volume in comparison to the constrained posterior. 

\begin{figure}[!ht]
    \centering
    \includegraphics[scale=0.34]{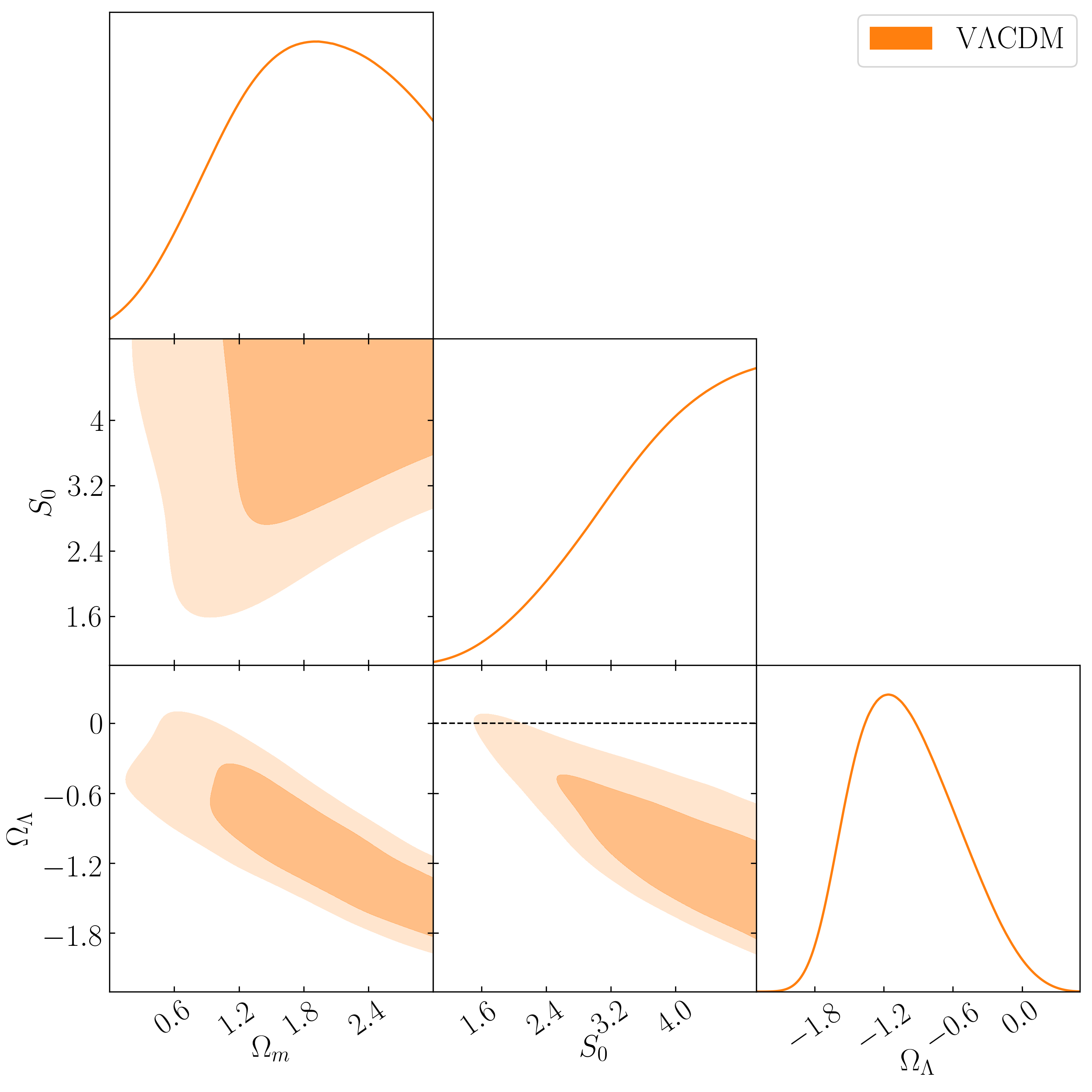}
    \caption{Confidence regions for the generalized  V$\Lambda$CDM model with three free parameters $\Omega_m$, $S_0$, $\Omega_\Lambda$, using CC+SN+GRB dataset. Here we show only the region implemented with the physical prior of $S_0>1$. Please note, that the apparent lower limit on the $\Omega_{\Lambda}$ is due to the assumed prior (upper limit on $S_0$).}
     \label{fig:VLCDM}
 \end{figure}

Although increasing the number of free parameters makes constraints/predictions less stringent, we discuss a more general model including three independent parameters $\Omega_m$, $\Omega_\Lambda$, $S_0$. In this case, we notice a multi-modal behavior of the likelihood determined by the prior regions of the parameters. The first one is, in fact, equivalent to $\overline{\Lambda{\rm CDM}}$ in which $S_0 \lesssim 1$ is either small or negative (ignoring for the moment that from the theoretical point of view $S_0$ has to be higher than unity). The second part of the parameter space corresponds to the VFD model extended by the cosmological constant term (V$\Lambda$CDM). 
For the V$\Lambda$CDM model the early-coasting and late-acceleration behavior of the VFD model is preserved and arrives at similar expectations for the quantities $\mathcal{O}_m(a)$ and $q(a)$, in fact, with the tighter constraint of $q(a=1)= -0.76\pm 0.10$. While this tighter constraint is contrary to intuition due to the additional parameter, it shows that the freedom to have a negative cosmological constant further aids the assertion of late-time acceleration. In this context, it is interesting to note that a very similar phenomenology based on dynamical symmetry breaking (see also \cite{Mannheim08}) was recently presented as a modified Friedmann's cosmology in \cite{Lu19}.
As is shown in \Cref{fig:VLCDM}, the theoretical prior of $S_0>1$ naturally gives rise to a negative value of $\Omega_\Lambda$ (i.e., a negative cosmological constant) and a highly degenerate scenario. Better calibration of $\Omega_m$, e.g., aforementioned primordial nucleosynthesis, would break the degeneracy providing the upper and lower bounds on $S_0$ and $\Omega_{\Lambda}$, respectively.

The low-redshift evidence for a negative cosmological constant was recently investigated in \cite{Visinelli19}, finding motivation in the fact that string theory might not accommodate stable de Sitter (dS) vacuum \cite{Vafa05,Danielsson2018} (see also \cite{Palti19}). Here a negative cosmological constant ($\Omega_{\rm cc} < 0$) was modeled alongside a dark energy component ($\Omega_{\phi}$) with EoS of $w_{\rm \phi} \neq -1$. Having no detection of the same, a lower limit of $\Omega_{\rm cc} \gtrsim -14.$ at $95\%$ C.L. was reported. However, in our modeling, we find a contrasting $95\%$ C.L. upper limit of $\Omega_{\Lambda} \lesssim -0.14$, while a lower limit remains degenerate with larger values of $S_0$ and $\Omega_m$. This extension also modifies the limits on the UV cutoff to $k_{max} > 12.9 [M_p/\sqrt{2+N_{sc}}]$ at $95\%$ C.L., which differs mildly but remains in agreement with the constraints set without the inclusion of a cosmological constant. It is worth stressing that the formalism considered is entirely different from that of \cite{Visinelli19}, and that $\Lambda$CDM is not a part of the allowed parameter space in our analysis, due to the prior $S_0>1$. The better-fits to the data are shifted towards the higher limits of the priors, as reported in the last column of \cref{Tab:priors}. We also find that this extension does not improve the AICc statistics over the VFD model within the assumed prior ranges, being disfavored w.r.t. $\Lambda$CDM at $\Delta {\rm  AIC} \sim 4.86$. Incidentally, this is also very similar to the $\Delta {\rm AIC} \sim 4.8$ reported in \cite{Visinelli19}, while using different datasets. However, using the Bayesian evidence, we find that the V$\Lambda$CDM model is very marginally disfavored with a $\ln\mathcal{B} = -1.9$, even with the very large prior volume available. That is indeed a good indication that further tests are very much desirable for the model in future considerations.

As a side note, we notice that the lower bound on $\Omega_{\rm cc}$ set in \cite{Visinelli19}, is essentially accompanied by $w_{de} \xrightarrow{} {-1} $ (consistent with $w_{\phi} =-1$ at $1\sigma$) and the marginalized posterior is limited by the assumed lower limit of the prior range. This makes the parameter space equivalent to $\Lambda$CDM and in-turn speculate the lower bound, as they model $\Omega_{\phi} \equiv 1 - \Omega_{\rm m} -\Omega_{\rm cc}  \xrightarrow{w_{\phi}\rightarrow-1} \Omega_{\Lambda}$ and this would imply an extended linear degeneracy between $\Omega_{\phi \rightarrow \Lambda}$ (positive) and $\Omega_{\rm cc}$ (negative) for all lower values than the quoted limit, having the well-constrained combination $\Omega_{\phi}+\Omega_{cc}\sim 0.65$.

As was already stressed, we imply that the negative cosmological constant originates, as from the vacuum fluctuations due to the Pauli sum rules violation \cite{Visser18}, so from field condensates. In the string theory (more generally, M-theory \cite{Horava1996, DUFF1996, Banks1997} or the F-theory \cite{Vafa1996, anderson2019ftheory}), only the last contribution, i.e., a false vacuum is usually considered \cite{Kallosh2015, Danielsson2018, kachru2019sitter, Palti19}, which is analogous to the condensate arising in the Higgs mechanism of particle masses generation in the standard model \cite{Commins1983}.
In particular, the string compactification \cite{Fraiman2018, grimm2019asymptotic} or bounding the strings by branes \cite{bena2019faces} results in a ``landscape'' \cite{Kallosh2015} or ``swampland'' \cite{Vafa05, Akrami18, Kim2019, grimm2019asymptotic, blumenhagen2019quantum, Roupec2018, Raveri2019, Garg2019, Obied18, Ooguri19, Palti19, Agrawal18, Andriot18} of the effective field theories \cite{colgin2019testing,agrawal2019h0,anchordoqui2019h0,Uzawa2018,angus2019mathbfodd}. Regarding an explicit consideration of the vacuum quantum fluctuations within the string theory \cite{Fontanini2006}, it is in an infantile state to date,  because the theory is, in fact, the first-quantized one, whereas its second quantized version is not completely realised yet \cite{MORRIS1988222, Thorn1988hm, Kaku2012} (however, see also \cite{Berglund2019}). That prevents considering the zero-point fluctuations consisting of the creation and annihilation of the strings. Only after reducing to effective field theory, the zero-point energy could be taken into account.

Also, one has to note that in the context of the generalized framework considered here, a gauge violating version of the string theory should be developed. For instance, it could be not a single moving relativistic string, but a system of the relativistic strings coupled into a crystal-like lattice. The preferred reference frame will appear in which such a ``crystal" is at rest ``in tote''.  At the same time, all linear perturbations of this system have to manifest Lorentz invariance.

\section{Discussion}
\label{sec:Conclusions}

As the well-known physicist Dmitry Blokhintsev stated: ``Amount of the facts are always sufficient - a fantasy is lacking" \cite{Blokhintsev59}. After the invention of the string theory and the loop quantum gravity \cite{Casares2018}, one could hardly reproach physicists in the absence of fantasy. However, this phrase could be understood in the sense that all the existing experimental facts are to be taken into account. Here we have tried to consider the cosmological constant problem as an observational fact demonstrating that the null energy of the quantum field oscillators does not influence universe expansion rate, and this viewpoint requires extending the GR framework.

We have utilized the conventional astrophysical data: SN, GRB, and CC to constrain the scenarios based on contributions of the residual vacuum fluctuations  against the standard $\Lambda$CDM model. Within this framework two models have been considered: $\overline{\Lambda{\rm CDM}}$ (i.e., the ${\Lambda{\rm CDM}}$ model extended by radiation like a vacuum contribution, with absence of the relation $\Omega_m+\Omega_\Lambda\neq1$) and  the VFD one, both having two independent parameters. In comparison to the standard ${\Lambda{\rm CDM}}$ model, the extended $\overline{\Lambda{\rm CDM}}$ and VFD models are disfavored with $\Delta$AICc of 1.92 and 4.02, due to one extra parameter  and an additional $\Delta\chi^2\sim 2.1$ disadvantage, respectively. 

However, such a preference can be elusive because the standard model, in fact, ignores the aforementioned vacuum energy problem considering the vacuum energy as non-existing really. It is a well-known issue that, at a given degree of accuracy of the observational data, a certain simpler model without a theoretical background allows describing the observational data better than a more fundamental but complex model. The considered $\overline{\Lambda{\rm CDM}}$-model shows that, after compensating for the main part of vacuum energy, the gravity framework could still be close to GR. In contrast, the VFD manifests a non-GR framework explicitly. Within this framework, we constrain the UV cut-off scale (\Cref{eqn:s0}) to be $k_{max} = 12.43^{+0.9}_{-1.6} [M_p/\sqrt{2+N_{sc}}]$, where $N_{sc}$ is a number of the minimally coupled scalar fields.

The consideration of the V$\Lambda$CDM model requires a negative value of the cosmological constant to agree with the astrophysical data at a $95\%$ C.L. upper limit of $\Omega_{\Lambda}<-0.14$. That contrasts with the lower limit reported in \cite{Visinelli19}, whose authors imply that the string theory insists on the negative cosmological constant, i.e., an AdS space arises from compactification \cite{ Danielsson2018}.In addition to the AIC we have also implemented the Bayesian evidence, where a similar inference is also made for the model-selection. While the VFD model is now strongly disfavored w.r.t. the $\Lambda$CDM model, having $\ln\mathcal{B} = -8.0$, the more general V$\Lambda$CDM extension allowing the negative $\Omega_{\Lambda}$ is only marginally disfavored with $\ln\mathcal{B} = -1.9$.

It should be reminded that the string theory considers the problem classically and arrives at the GR framework after compactification. Above, we have suggested a more general context (\ref{fried}) than GR, and considered the quantum effects of the different types, so that the obtained limit of $\Omega_{\Lambda}<-0.14$ is a limit to the overall cosmological constant consisting both of condensates and vacuum fluctuations.  It is expected that the string theory should also be able to account not only for the "landscape" \cite{Kallosh2015} false vacuum, but for the zero-point fluctuations over this landscape. 

Aside from the aforementioned physical motivations, leading to the current work, there exist several shortcomings and notable deviations from the $\Lambda$CDM, such as the $H_0$ tension \cite{Riess19,Wong19}, dark energy considerations \cite{wang18, Zhao17_nature} and curvature arguments \cite{Barrow11} (recently, \cite{Handley19, DiValentino19}). We intend to address some of these issues through the generalized framework presented here, in future communication, also with an anticipation that the interpretation of high-$z$ observables, such as CMB will be taken into account. 

\section{Acknowledgements}
B.S.H acknowledges financial support by ASI Grant No. 2016-24-H.0. B.S.H acknowledges INFN Roma, Tor Vergata Computing Centre services (RMLab) and is thankful to Federico Zani for providing help with the same. B.S.H is thankful to Maurizio Firrotta for useful discussions and we thank Massimo Bianchi for constructive comments on the draft. V.L.K. acknowledges the support by the Marie S.-Curie Cofund Multiply ``MASTEDIS'' Fellowship (grant No. 713694).

\bibliography{bibliografia}

\end{document}